\documentstyle[aaspp4,12pt,epsfig,color]{article}
\def\simlt{\lower.5ex\hbox{$\; \buildrel < \over \sim \;$}}

\def\stacksymbols #1#2#3#4{\def\theguybelow{#2}
\def\verticalposition{\lower#3pt}
\def\spacingwithinsymbol{\baselineskip0pt\lineskip#4pt}
\mathrel{\mathpalette\intermediary#1}}
\def\intermediary#1#2{\verticalposition\vbox{\spacingwithinsymbol
\everycr={}\tabskip0pt
\halign{$\mathsurround0pt#1\hfil##\hfil$\crcr#2\crcr
\theguybelow\crcr}}}

\def\lapproxeq{\stacksymbols{<}{\sim}{4}{1}}

\begin{document}

\title{THE FORMATION OF GALAXY DISKS}

\author {Joseph Silk}
\affil{Astrophysics, Department of Physics, University of Oxford, Oxford OX1
3RH UK; silk@astro.ox.ac.uk}

\begin{abstract} 
Galaxy disk formation must incorporate 
the multiphase nature of the interstellar medium. The resulting
two-phase structure is generated and maintained 
by gravitational instability and supernova energy input, which
yield  a  source of turbulent viscosity
that  is able to effectively compete  in the protodisk phase
with   early angular momentum loss of the
baryonic component 
via dynamical friction in the dark halo. Provided that star formation occurs on the viscous drag time-scale,  this mechanism 
 provides a means of 
accounting for disk sizes and radial profiles. 
 The star formation  feedback  is self-regulated  by  turbulent gas
pressure-limited
percolation of the supernova remnant-heated hot phase,
but  can run away in gas-rich protodisks to
generate compact starbursts.
A simple analytic model is  derived for a Schmidt-like global star
formation law in terms of 
the  cold gas volume density.

\end{abstract}
{\it Subject headings:} galaxies:formation--galaxies: ISM--galaxies:
kinematics and dynamics-galaxies: stellar content

\section{Introduction}

 An outstanding problem in understanding the formation of galaxy disks
is the origin of the characteristic disk scale length. In the absence of
a theory that determines this parameter, one can place little credence
in simulations of numerical or semi-analytical galaxy formation. Yet
until recently, the elements of a robust theory for disk scales appeared
to be in place.

Tidal torques between neighbouring density fluctuations generate an
initial dimensionless angular momentum $\lambda_i \equiv J \vert E \vert
G^{-
1}M^{-5/2} \approx 0.05$, with a broad dispersion $\Delta \lambda_i \sim
\lambda_i$. Baryons cool and contract once dark halos virialize, conserving
specific angular momentum to form a centrifugally supported disk when
$\lambda
\sim 1$. The final disk radius is $r_d \sim \lambda_i r_h$, where $r_h$
is the halo radius. This yields a disk scale length of several kpc for
an $L_*$ galaxy, as observed (Fall and Efstathiou 1980).

However with the advent of numerical hydrodynamical simulations with
adequate resolution, it has become apparent that angular momentum is
effectively transferred outwards as dense baryon clumps sink into the
central regions of the forming galaxy. Clumpiness in the dark halo also
tidally perturbs the disk and aggravates the angular momentum transfer
problem. The result of the simulations is that the disk sizes are
invariably too small (Steinmetz and Navarro 1999), by a factor  $\sim$ 5.

Two schemes have been proposed in order to halt excessive
angular momentum transfer. Photoionization of the intergalactic medium
destroys small-scale substructure (Navarro and Steinmetz 1997). These
structures are the building
blocks of large-scale structure and a source of clumpiness during
massive halo formation. However the CDM structure is unaffected and
there is little effect on the formation of massive galaxies. Another
problem related to that of angular momentum transfer in that it involves
baryonic dissipation is that of overcooling. This results in premature
exhaustion of the gas supply. A solution designed to tackle both of these
problems appeals to feedback from star formation via supernovae.
Provided gas cooling is delayed to $z \sim 1$, one can form disks of the
appropriate size and mass (Weil, Eke and Efstathiou 1998).

However this success comes at a price. Late disk formation implies
strong evolution in disk size, luminosity and surface brightness at $z
\sim 1$. Large disks are observed to $z \sim 1$, and there is little
evidence for any evolution in scale length and  surface brightness, other
than passive stellar population ageing (Lilly et al. 1998). Disks are
observed to $z\sim 1$ with a Tully-Fisher relation  that is essentially
indistinguishable from that of present-day disks, apart from passive
evolution.  Not all analyses agree: Bouwens and Silk (2000) find that up to
2 mag of surface brightness evolution is required at $z\sim 1$ in an
analysis of a much larger data set
(Simard et al. 1999). However the very existence of substantial numbers of
disk galaxies at  $z\sim 1$  means that 
early heating of protodisks  and consequent suppression of star formation to
$z\sim 1$ appears to be an unacceptable resolution of the
angular momentum transfer problem.

A new approach is required. I have previously proposed  (Silk 1997)
the following model for disk
formation. Feedback from supernovae results in a two-phase interstellar
medium,
the porosity of which controls  feedback on the supply and 
compression of cold gas and the ensuing star formation rate.
Gas-rich disk gravitational instabilities drive cloud-cloud collisions
that 
provide  an effective turbulent 
viscosity which is responsible for the formation and
contraction of the disk. Viscous galactic
disk models have a long history, commencing with the original
proposal by Silk and Norman (1981),
the  definitive study by Lin and Pringle (1987a) that derived exponential
stellar surface density profiles, and the chemical evolution studies of Clarke (1989,1991).
In the cosmological context of cold dark matter halos,
turbulent viscosity has 
 been incorporated into  disk models by Firmani et al. (1996). I show here
that
the
 viscous drag time-scale is of
the same order as the dynamical friction time-scale that controls the loss 
of angular momentum from baryons to dark matter (Section 2). 
I apply porosity feedback to derive a Schmidt-like star formation law that
depends on cold gas
volume density (Section 3), and  I examine outflows and radial flows (Section 4).
The requirement
that the 
star formation time-scale is on the order of the viscous time-scale suffices
to give an exponential disk with an appropriately large-scale length and that moreover
forms at a relatively early stage (Section 5).

\section{Angular momentum problem resolved}

The loss of baryonic angular momentum during disk formation is primarily due
to the dynamical friction
exerted by the inner dark halo on the baryonic clumps. The time-scale for
dynamical friction to act
is approximately $(M/M_{cloud})$ dynamical times, or
$$ t_{df} \approx \left( \frac{r}{V} \right) {\left( \frac{r}{l_t} \right)
}^3 \approx \frac{r}{V} {\left(
 \frac{V}{\sigma_g} \right)}^3$$
\noindent where the first equality made use of the tidal radius $l_t$
defined by
$$ \rho / \bar{\rho} = ( l_t/l)^3$$
for a cloud of density $\rho$ and size $l$ in a medium of mean density
$\bar{\rho}$. The second equality
replaces the disk circular velocity $V$ by $\Omega r$ and $\sigma_g$ by
$\Omega l_t$, the latter being
the cloud velocity dispersion induced by gravitational instabilities (Gammie,
Ostriker and Jog 1991).

To compare $t_{df}$ with the viscous drag time-scale, defined by
$$t_{\nu} = r^2/ \nu,$$
I compare three alternate definitions of viscosity. In a non-self-gravitating cold accretion
disk, the $\alpha$-prescription (Shakura and Sunyaev 1973) is commonly used:
$$ \nu_1 = \alpha H c_s \qquad \alpha \sim 1,$$
where $H$ is the scale height and $c_s$ is the sound velocity. This
prescription  is lacking in 
fundamental motivation. A recent proposal
(Duschl, Strittmatter and Biermann 2000) appeals
to turbulence experiments which
suggest that a limiting
Reynolds number
$R_{crit} \sim 10^2 - 10^3$ demarcates the onset of turbulence, and yields
the ansatz
$$ \nu_2 = \beta r V, \qquad \beta \approx R^{-1}_{crit}.$$
This expression should be
applicable to self-gravitating systems and to
spheroidal as well as disk geometries. Finally, 
non-axisymmetric  
gravitational instabiliies of
a cold disk provide an effective turbulent viscosity that transfers
 angular momentum (Lin and Pringle 1987b).
This results in a characteristic
instability scale $(l_t)$ and turbulent velocity $(\sigma_{g}\approx
\Omega l_t)$,
motivating the hypothesis that 
$$
\nu_3 = \gamma l_t \sigma_{g}, \qquad \gamma \sim 1.$$
The corresponding time-scales for viscous drag are 
$$t_{\nu_1} = \frac{1}{\alpha} \left( \frac{r}{V} \right) \left(
\frac{\sigma_g}{c_s} \right) \left( \frac{V}{\sigma_g} \right)^3; \qquad
t_{\nu_2} = \beta^{-1} \left( \frac{r}{V} \right); \qquad t_{\nu_3} =
\frac{1}{\gamma} \frac{r}{V} \left( \frac{V}{\sigma_g}
\right)^2.$$
In all three prescriptions for viscosity, the viscous time is of order
the dynamical friction time; and indeed $t_{\nu_3}$ (and even
$t_{\nu_2}$) can be substantially less. It is likely that only
 $t_{\nu_2}$ and $t_{\nu_3}$
are relevant to self-gravitating disk formation, and I conclude that
viscosity in the
protodisk can operate more effectively than dynamical friction.

This conclusion would not resolve,
and indeed would even aggravate,  the angular momentum problem, since
viscous drag also removes angular momentum from the gas clouds, were it
not for the fact that the star formation time $t_*$ itself likely to be
comparable to the viscous time. This is plausible because the radial gas
flow $\dot{M}_g = 2 \pi \Sigma_g \nu$ is regulated by viscosity, and the gas
flow controls the gas supply  available for star formation at any
given radius. Strong support for this conjecture comes from the model of a
self-gravitating viscous disk, in which Lin and Pringle (1987a)
find that $t_{\nu} \approx t_*$ results in an exponential surface
density profile.
Hence I conclude that star formation develops before substantial angular
momentum loss is induced by dynamical friction of infalling clouds against
the dark matter, and that the resulting disk scale-length is therefore
of order $\lambda_i r_h$, where $r_h$ is the initial halo radius $( \sim
100$ kpc) at overdensity 200 (the virialization radius) and $\lambda_i
\approx 0.05$ is the dimensionless halo angular momentum parameter
induced by gravitational torques in the early universe. 
Moreover the resulting stellar disk surface density profile is exponential. Of course,
numerical
simulations are needed to verify these conjectures and should eventually be
feasible.

\section{Global star formation rate: the second parameter}

It is usually assumed that the star formation rate in a self-gravitating
gas-rich disk is regulated by gravitational instability according to
whether  the Toomre parameter 
$$Q_g \approx \frac{ \delta \kappa \sigma_g}{\pi G \Sigma_g}$$ 
is smaller or larger than unity, for instability or stability
respectively. Here, $\kappa$ is the epicyclic frequency, $\sigma_g$ is the
gas-velocity dispersion, $\Sigma_g$ is the gas surface density, and
$\delta$ is a parameter, of order unity for a gas-rich disk, that allows
for the stellar contribution to the self-gravity of the gas component of the
disk. H$\alpha$,
HI and CO mapping of disks shows that $Q_g \sim 1$ in the star-forming regions
of disk
galaxies, suggesting that star formation is self-regulated by
gravitational instability (Kennicutt 1989). This is plausible, since disk instabilities
heat the disk and raise $Q_g$ via tidal acceleration of clouds, whereas gas
cloud collisional dissipation decreases the gas velocity dispersion,
decreasing $Q_g$. Star formation occurs in the clouds that form by
gravitational instability, grow in mass by coalescence, and become
unstable to fragmentation. $Q_g$ self-regulation predicts that there is a
threshold to star formation. Star formation ceases when the gas surface
density drops below a critical value
$$\Sigma_{crit} \approx \delta \kappa \sigma_g/ \pi G.$$
Now $\kappa \approx 2^{\frac{1}{2}} \Omega$ for a flat rotation curve,
$\Omega \approx 10^{-15}$ s$^{-1}$ and $\sigma_g \sim 10$ kms$^{-1}$
for the Milky Way, as a typical example, so that $\Sigma_{crit}
\approx 10 M_{\odot}\rm  p c^{-2}$. Clouds of giant HII regions, both in our
galaxy and in nearby spirals, support this star formation threshold
(Kennicutt 1998). However H${\alpha}$ mapping (Ferguson et al. 1998) has
shown
 that
substantial star formation occurs at $\Sigma < \Sigma_{crit}$. To
properly account for star formation, a second parameter is evidently
needed that supplements or modifies the $Q_g$ criterion.

A promising direction for such a modification would involve the
dependence of the star formation rate on gas volume  density, rather than surface
density. Disk flaring develops in the outer disk, and the resulting density
decrease at fixed $\Sigma_g$ would be a logical inhibitor of star
formation that occurs near, but independently of,   the radius at which
$\Sigma_g$ drops below $\Sigma_{crit}$. Indeed, it is entirely possible
that the physics that controls disk flaring, namely the transition between
the dominance of disk and halo self-gravity, is ultimately responsible
for determining the transition between the $Q_g < 1$ (inner disk) and $Q_g >
1$ (outer disk) regimes. I develop here a physical criterion for
controlling the star formation rate via feedback that depends primarily
on the gas density, rather than the gas surface density. Supernovae
control the energy and momentum input that heats and stirs the interstellar
medium. The rate of supernovae is controlled by the star formation rate,
and therefore provides a natural means of self-regulating star formation.
The essence of this self-regulation can be described by a two-phase
model of the interstellar medium, consisting of the hot gas directly
heated by supernova remnants and the cold gas that is compressed by
supernova blast waves into dense shells,  dense cold clumps of which
also survive the passage of the supernova blast waves.

 I have argued  that
global feedback operates via the porosity of a two-phase interstellar
medium (Silk 1997). I now further extend these arguments by deriving an
expression for the star formation rate that does not explicitly depend
on $Q.$
The porosity is related to the hot gas fraction by $P =
-\ln(1-f)$,
where $f$ is the volume filling factor of hot gas. For global feedback, the
hot gas volume fraction
must be significant, say $f > 0.1$. If the gas density or pressure
were sufficiently high, supernova feedback would be restricted to such
small volumes that there would be no global effect. A two-phase
interstellar medium is adopted in order to describe the interaction of
cold and hot gas. The cold gas is atomic and molecular gas, and
initially comprises the entire disk. Once stars form and die, a steady
state is reached in which bubbles of hot gas driven by supernovae sweep
out and envelop much of the cold gas. The hot gas mass fraction is
determined by cold cloud evaporation and by cooling of the hot gas.
The diffuse hot gas erodes cold clouds and  leaks out of the disk via
chimneys and fountains. I will estimate  below the rate at which gas ejection
occurs.
The porosity is a convenient and calculable parameter that  defines the
filling factor of the hot phase, and hence its ability to diminish the
efectiveness of star formation by eroding the cold gas supply.

Supernova remnants  deposit
energy and, ultimately,  momentum into the interstellar gas at a rate
$\dot{\rho_*} v_{SN}$ where
$v_{SN}$ is the specific momentum injected by supernovae per unit star
formation rate $\dot\rho_*$ and is given by 
\begin{equation}
v_{SN} = E_{SN} /v_c m_{SN} \ = 500 E_{51}^{ 13/14} n_g^{- 1/7} m_{250}
\zeta_g^{-3/14}  \rm km\,s^{-1} \label{eq:17}
\end{equation}
with $v_c = 413 E^{1/4}_{51} n^{1/7}_g
\zeta^{3/14}_g$ being the velocity at which the remnant enters
the momentum conserving regime, $E_{51} \equiv E_{SN} / 10^{51}$ergs the
supernova energy (taken to be $10^{51}$ ergs), and $\zeta_g$ the
metallicity relative to solar of the ambient gas (Cioffi, Mckee and
Bertschinger 1988). Also $m_{SN}$ is the
mean mass required in forming stars in order to produce a supernova. For
SNII,
one simply assumes an initial mass function (IMF) with all stars of mass
above 8 M$_{\odot}$ becoming supernovae, so that $m_{SN} \approx 250$
M$_{\odot}$ 
for a Miller-Scalo IMF. One can crudely double the inferred rate
for an estimate of the rate of Types I and II supernovae after the first
$10^8$ years have elapsed (to allow sufficient time for SNIa to form).

Porosity is defined to be the product of the supernova remnant 4-volume at
maximum extent, when halted by ambient gas pressure, and the
rate of bubble production. 
 The porosity is a measure of
the fraction of volume $f$ occupied by the hot phase ($T
\sim 10^6\,{\rm K}$) associated with the interiors of
supernova remnants.  By definition $f=1-e^{-P}$, and the
porosity $P=(\dot\rho_* /m_{SN})
\left(\frac{4}{3} \pi\, R^3_a\, t_a \right)$
where $\dot\rho_*$ is the star formation rate, $m_{SN}$ is
the mass in stars formed per supernova, and $R_a$ is the
radius of the supernova remnant at time $t_a$ when halted by
the ambient (turbulent) gas pressure $p_g$.  One finds that
the porosity $$ P \propto \dot\rho_*\, p^{-1.36}_g \rho^{-0.11}_g $$
is extremely sensitive to the interstellar pressure.  This
 provides the motivation for the feedback
prescription.
By incorporating an analytic fit to the
evolution of a spherically symmetric supernova-driven shell (Cioffi et al.  1988), one can
write
\begin{equation}
P = G^{-\frac{1}{2}} \sigma^{2.72}_f p^{-1.36}_g \rho^{-0.11}_g
\dot{\rho}_* \ , \label{eq:19}
\end{equation}
where $\dot{\rho}_*$ is the star formation rate, $p_g$ is the ambient
gas pressure, both thermal and turbulent, and $\sigma_f$ is a fiducial
velocity dispersion that is proportional to $E^{1.27}_{SN} m^{-1}_{SN}
\zeta^{-0.2}_g$ and may be taken to be 22 km s${^{-1}}$ for $E_{SN} =
10^{51}$ erg, $m_{SN} = 250$ M$_{\odot}$ and $\zeta_g = 1$. Note that at
large pressure, $P \ll 1$, and porosity is primarily controlled by the ambient
pressure, which I take to be dominated by turbulence: $p_g = \rho_g
\sigma^2_g$.

 Feedback is regulated by the porosity parameter $P$.
 Previously, the star formation rate had no explicit feedback. Consider the
possibility of strong feedback. In this case,  the filling factor of hot
gas is of order fifty percent, which is the requirement for strong feedback.
Rewriting (2) as 
\begin{equation}
\dot{\rho}_* \approx G^{1/2} \rho^{3/2}_g \left(
\frac{\sigma_g}{\sigma_f} \right)^{2.7} P \ ,  \label{eq:20}
\end{equation}
one explicitly incorporates feedback into the star formation rate.
The natural limit in which
feedback self-regulates has $P \sim f \sim 1/2$. 
If $P\ll 1,$ feedback is irrelevant: if  $P\gg 1,$ star formation would
be completely suppressed. It seems likely, though hard to prove rigorously, that $f\sim 1/2$ 
 corresponds to self-regulation of 
disk star formation. Certainly, $f\sim 0.2-0.5$ 
 prevails in the local interstellar medium (Slavin  and Cox 1993; Ferriere 1995). 
In this case, the
star formation rate is independent of $Q$, and is controlled by
ambient pressure. The turbulent pressure is fed by disk gravitational
instabilities, and  implicitly
depends on $Q.$
The derived global star formation law has a normalisation, equivalent to
star formation efficiency, that depends explicitly on cloud velocity
dispersion.
 The star formation rate  and efficiency  are 
enhanced by stirring: this might be expected near a bar or as a consequence
of a merger.

Another limiting case is $P \sim f < < 1$. Here porosity is small and
feedback is unimportant. This might be relevant in a dense gaseous
protodisk, and would suggest the possibility of runaway star formation.

\section{Feedback and outflow}

Supernovae are responsible for the feedback from star formation that
renders star formation inefficient. The star formation efficiency
$\varepsilon$ may be defined explicitly by writing
the star formation rate as 
$$ \dot{\rho}_* = \varepsilon \Omega \rho_g,$$
where $\varepsilon$ is the fraction of gas converted into star per dynamical
time. Adapting
the porosity model for feedback, one now can write the porosity as
$$ P \approx \left( \frac{\rho}{\rho_g} \right)^{\frac{1}{2}} \left(
\frac{\sigma_f}{\sigma_g} \right)^{2.7} \varepsilon,$$
where $\rho$ is the total (gas plus star) density in the disk or protodisk.
High star formation efficiency drives high porosity and significant
feedback. 

The momentum input from supernovae is dissipated via cloud-cloud
collisions and outflow from the disk. In a steady state, the momentum
input rate $\dot{\rho}_* v_{SN}$ must balance the cloud collisional
dissipational rate $ \rho_g \sigma^2_g l^{-1}_t$ and the momentum
carried out in outflows $f p_g H^{-1}$, where $p_g$ is the turbulent
pressure $\rho_g \sigma_g^2$ of the two-phase interstellar medium, $H
\equiv \sigma^2_g/2 \pi G \Sigma$ is the disk gas scale height, and $\Sigma$
is the surface mass density. At a given disk radus, there is also
viscosity-driven inflow, $M_g = 2 \pi \Sigma_g \nu$, which
provides a gas supply for star formation. Ignoring the outflow and
infall contributions in the momentum budget, one has 
$$\varepsilon \Omega \mu_g v_{SN} = \mu_g \sigma_g \Omega,$$
so that $\varepsilon = 0.02$ $(\sigma_g / 10$ kms$^{-1}$) $(500$ kms$^{-
1} / v_{SN})$. This suggests that supernova feedback can indeed yield
the required low efficiency of star formation.

For a galaxy such as the Milky Way, the  global
star formation efficiency is expected to be around 2 percent, both as inferred from the global
values of gas mass $( \sim 6 \times 10^9 M_{\odot})$ and star formation
rate $( \sim 3 M_{\odot} $yr$^{-1}$) after allowance for gas return from
evolving stars (the returned fraction $\sim 0.5$ for a Miller-Scalo IMF)
over a galactic dynamical time and as more
directly inferred  from
studies of HII region radio luminosities summed over molecular cloud masses
 (e.g. Williams and McKee 1997).

Inserting the derived expression for star formation efficiency into the
equation for the porosity, I find that
$$ P \approx \left( \frac{\rho}{ \rho_g} \right)^{\frac{1}{2}} \left(
\frac{\sigma_f}{ \sigma_g} \right)^{1.7}
\frac{\sigma_f}{v_{SN}} = 0.5 \left( \frac{\rho / \rho_g}{0.1}
\right)^{\frac{1}{2}}
\left( \frac{\sigma_g}{10 \rm{kms}^{-1}} \right)^{- 1.7}.$$
The observed three-dimensional cloud velocity dispersion is 11 kms$^{-
1}$ (for molecular clouds within 3 kpc of the sun) (Stark and Brand 1989). Thus the Milky Way
interstellar medium has predicted porosity comparable to what is
observed.

If the hot volume fraction is large, outflow from the disk is
inevitable. The ratio of outflow to star formation rate per unit area of
disk be written as ($f_s p_g H^{-1} V_{esc}^{-1}) ( \varepsilon \Omega
\rho_g)^{-1}$, where  $V_{esc}$ is the escape velocity from the disk.
Now $p_g = \pi G \Sigma_g \sigma_g = \frac{1}{2} \Sigma_g
\Omega V$ and $\Sigma_g = 2 \rho_g H$, so that the outflow rate
simplifies to 
$$ \frac{\rm \ outflow \ rate}{\rm star \ formation \ rate} \approx
\frac{f_s}{\varepsilon},$$
where $f_s$ is the surface area covered in hot bubbles.

This ratio will typically be, for say $f \sim P \sim 0.3$ and $ \varepsilon
\sim 0.03$, around 10, and is comparable to what is observed for both
star-forming dwarfs and more luminous disk galaxies (Martin 1999). In
general, the bulk of the gas cannot escape in a wind from normal galaxies, otherwise
the gas reservoir would be seriously depleted. Known infall rates amount
to at most 10 percent of the star formation rates. The gas must cool in
the halo and fall back into the disk.

The viscosity prescription leads to a radial inflow of the gas. This
amounts to $v_r = \nu r^{-1}$, which can be rewritten as
$$v_r = \gamma \sigma_g l_t r^{-1} = \gamma \sigma^2_g / V,$$
where I have used the expression for $\sigma_g$ derived for
gravitational instability-driven random motions. Hence one typically
expects
$$v_r \approx 0.5 \gamma \left( \frac{\sigma_g}{10 {\rm km\,s}^{-1}}
\right)^2
 \left( \frac{{200\rm km\,s}^{-1}}{V} \right) {\rm km\,s}^{-
1}.$$
The associated mass flux is
$$\dot{M}_g = 2 \pi \nu \Sigma_g = 2 \pi \gamma l_t \sigma_g \Sigma_g.$$
Again making use of $\sigma_g = \Omega l_t$ and $\Sigma = \Omega V
/ 2 \pi G$, the mass flux reduces to 
$$ \dot{M}_g = \gamma \sigma^2_g V G^{-1} ( \Sigma_g/ \Sigma) =
\gamma( \sigma_g / V)^2 \Omega M ( \Sigma_g / \Sigma).$$

Comparing this with the global star formation rate $\dot{M}_* =
\varepsilon \Omega M_g$, with $ \varepsilon = \sigma_g / v_{SN}$,
one obtains
$$\dot{M}_g = \gamma ( \sigma_g / V) ( v_{SN} / V )
\dot{M}_*,$$
or
$$ \dot{M}_g / \dot{M}_* = \left( \frac{ \gamma}{8} \right) \left(
\frac{\sigma_g}{10 {\rm kms}^{-1}} \right) \left( \frac{v_{SN}}{500 {\rm
kms}^{-1}} \right) \left( \frac{200 {\rm kms}^{-1}}{V} \right)^2.$$
I conclude that the viscous supply of gas may account for of order 30
percent of the net gas supply, after return, as is required to balance star
formation. The associated radial flow
 is sufficient to modify chemical evolution
in the solar neighbourhood and resolve the G dwarf problem (Clarke 1991),
 and to generate a disk metallicity gradient (Clarke 1989).
However the most novel application is to the disk scale length, as I now
demonstrate.

\section{Disk properties}

The disk size may be estimated by setting the viscous time-scale equal
to the star formation timescale. This guarantees an exponential
profile, and the characteristic length scale is obtained by setting $t_{\nu} =
t_{sfr}$, so that 
$$ r_{d,\nu} = \gamma \sigma^2_g t_{sfr} / V.$$
Inserting the star formation time-scale
$$t_{sfr} = \frac{1}{\varepsilon \Omega} \frac{\rho}{\rho_g} =
\frac{1}{\Omega} \left( \frac{\rho}{\rho_g} \right) \left(
\frac{v_{SN}}{ \sigma_g} \right),$$
I infer that
$$
r_{d,\nu} = 
\left( \frac{\gamma v_{SN}\sigma_g}{2\pi G\Sigma} \right) 
 \left( \frac{\rho}{\rho_g} \right)\\
= 2 \gamma \left( \frac{v_{SN}}{500 {\rm
kms}^{-1}} \right) 
\left( \frac{\sigma_g}{10 {\rm kms}^{-1}} \right) 
\left( \frac{50 \rm
M_\odot\,pc^{-2}}{\Sigma} \right) \left( \frac{\rho}{2 \rho_g} \right) {\rm
kpc},
$$
comparable to the scale lengths of Milky Way-type disks.

It is interesting to compare this with the standard derivation of the
final radius of a nonlinear cold dark matter density perturbation that
collapses to approximate virialization at a spherically-averaged  
overdensity of 200. The radius at mean overdensity 200 is
$$ r_{200} \approx \frac{0.01V}{H(z) \Omega_0^{\frac{1}{2}} } ,$$
where $V$ is again the circular velocity, $H(z)$ is the Hubble parameter, and
$\Omega_0$ is the density parameter. The baryons dissipate and contract
within the
cold dark matter halo to a final disk radius, if angular momentum is
conserved, of 
$$r_{d,cdm} = \lambda_i r_{200}$$
In the viscous model, the disk radius can be written in terms of the
virial radius defined by the self-gravitating baryonic component,
$r_v = {V^2}({ \pi G \Sigma})^{-1},$ so that
$$
r_{d,\nu} = \frac{\gamma}{2} \left( \frac{v_{SN} \sigma_g}{V^2} \right)
\frac{\rho}{\rho_g} r_v.
$$
Allowing for the initial baryon fraction $f_b$ (expected to be $\Omega_b
/ \Omega_m \sim 0.05 / 0.3 \sim 1/6)$, one finds that the gas first becomes
self-gravitating at radius $ f_b r_{200}.$ 
Hence the disk radius is comparable to that attained in the
idealized cold dark model, with conservation
of specific angular momentum of the baryons,
$$\frac{r_{d,\nu}}{r_{d,cdm}} = \frac{f_b}{\lambda_i} \frac{\gamma}{2} \left(
\frac{v_{SN} \sigma_g}{V^2} \right) \left( \frac{\rho}{\rho_g} \right) \sim
1.$$

The strong dependence of $r_v$ (or $r_{200}$) on $z$  leads to the
predicted sensitivity of the Tully-Fisher relation to redshift (Mao, Mo and
White 1998), in
contradiction with observations (Voigt 1999). In the viscous model, this
effect
is diluted because of the proportionality of disk radius to gas velocity
dispersion.
I expect $\sigma_g$ to increase with redshift because the effect of mergers
and
infall will result in systematically more turbulent protodisks. Hence this
should
partially compensate the reduction of $r_d$ at earlier epochs arising  from
the 
explicit redshift dependence in the expression for $r_{200}$.
Some reduction in$r_d$ is neded in order to account for the observed
evolution in disk surface brightness at $z\sim1$ (Bouwens and Silk 2000).

The Toomre parameter for disk instability may be rewritten as 
$$Q_g \approx \frac{\delta \Omega \sigma_g}{\pi G \Sigma_g} = 2 \delta \left(
\frac{\Sigma}{\Sigma_g} \right) \left( \frac{\sigma_g}{V}
\right),$$
where $\Sigma$ is the total disk surface density of stars plus gas.
Initially $Q_g <1$, so that the protodisk was unstable. At late times
however, $Q_g$ increases as $\Sigma_g$ decreases, although $\sigma_g$
also decreases, so this increase in $Q_g$ is partially compensated by the
decrease in gas turbulence, and the disk should remain
at least marginally
unstable, $Q_g \lapproxeq 1.$

To maintain $Q_g \sim 1$ over many dynamical times, as required for disks of
spiral galaxies, the disk must remain gas-rich. Gas cooling is required
to dissipate the dynamical instability heating. One can estimate that the
cold
gas controls the gravitational instability of the disk if $\mu_g/
\sigma_g > \mu_*/ \sigma_*$, and this condition requires a gas fraction
of 30 percent or more, if the gas velocity dispersion $\sigma_g \sim 10$
km s$^{-1}$ and the stellar velocity dispersion $\sim 30$ km s$^{-1}$.
This is indeed observed for the solar neighbourhood, where $\mu_g
\approx 15$ M$_{\odot} \ $pc$^{-2}$ and $\mu_* \approx 40$ M$_{\odot} \
$pc$^{-2}$.  The gas supply is prescribed by radial inflow in
the viscous  disk model, as is consistent with the gas reservoir 
inferred from observations of
the high velocity clouds.

\section{Discussion}

In summary, I have argued that the 
justification for the viscous disk model comes from the fact that
viscous
and star formation time-scales are likely to be comparable,
$$ \frac{t_{sfr}}{t_{\nu}} = \gamma \left( \frac{\rho}{\rho_g} \right)
\left(
\frac{\sigma_g}{V} \right) \left( \frac{v_{SN}}{\Omega_d r_d} \right),$$
where $\Omega_d$ is the rotation rate evaluated at the disk scale length. 
The baryons retain most of their initial angular momentum
since the 
dynamical friction time-scale exceeds the viscous time-scale,
$$ \frac{t_{df}}{t_{\nu}} = \gamma \left( \frac{ V}{\sigma_g} \right).$$
This consequently justifies an exponential disk profile  with  scale length of order
$r_{d,\nu} \approx f_b ( v_{SN} \sigma_g / V^2) r_{200}$. This scale 
coincidentally
happens to  be of order $\lambda_i r_{200}$, confirming that
approximate angular
momentum
conservation is achieved, and disk sizes match the observed range.

The viscous disk model predicts a radial inflow rate of order 10\% of the
disk star formation rate. This will generate a gradient in the chemical
abundances, and also provides a source of gas for bulge formation during
the phase of early gas-rich disk evolution. The predicted ratio of bulge
to disk stellar mass is of order $\sigma_g v_{SN} / V^2$, i.e. of order
10\%. Late forming bulges by early viscous disk evolution would imply
that bulges are more metal-rich than the disk, and that the age spread
of bulge stars is comparable to that of the old disk.

Supernova feedback regulates the star formation rate. The dominant parameter
is
porosity, largely determined by the turbulent gas pressure $p_g.$
 When $p_g$ is in
the range
expected for typical disk galaxies, $P \sim 1$ and feedback is important. I
assume
that the feedback is negative: further work is needed to
 justify this.
In the case of high turbulent pressure, expected in the aftermath of
mergers, $P \ll 1$. The
lack of feedback means that star formation can run away, only being limited by
the available
gas reservoir. In general, I find that supernova feedback results in a star
formation rate
$$ \dot{\rho}_* \approx \Omega \rho_g ( \sigma_g / v_{SN} ).$$
Thus even in starbursts, a Schmidt-type relation is maintained. The
predicted efficiency
can be as high as 20 \%, for merger-induced turbulence $\sigma_g \sim 100$
kms$^{-1}$ and a 
standard IMF, for which $v_{SN} \approx 500$ kms$^{-1}$. 

A top-heavy IMF
could substantially
reduce $v_{SN}$, and star formation efficiences of order 50\% would then
be attainable.
Such efficiencies may be needed in order to account
for the luminosities measured in many ultraluminous infrared galaxies where the
molecular gas masses are measured.  A top-heavy
stellar initial mass function might be required in protogalaxy mergers in
order to reconcile the hypothesis that ellipticals form in such events with
the observed paucity of young ellipticals at intermediate redshifts.

Gas density (and pressure) rather than surface density is the dominant
gas parameter in porosity-regulated star formation. This means that disk
flaring at the transition from disk self-gravity to halo gravity
dominance, where the gas scale height increases due to the reduction in
disk self-gravity, will tend to result in a reduction  of the disk star
formation rate. Disk gravitational instability, which controls the supply
of massive molecular clouds, is also quenched at approximately the same
radius. This leads to a more complex dependence of star formation rate
on gas surface density than suggested by the simple $Q_g$ threshold models
that have hitherto been advocated.

I thank J. Devriendt for discussions and a critical reading
of the manuscript. While this paper was in a refereeing phase 
that was protracted for reasons beyond my control, a paper appeared that contains 
an extensive discussion of feedback in a similar context to that discussed in the present paper (Efstathiou 2000). A more detailed comparison of the two approaches will be presented elsewhere.

\def\mnras{MNRAS}
\def\araa{ARAA}
\def\apj{ApJ}
\def\aj{AJ}
\def\pasp{PASP}
\def\apjl{ApJ}

\end{document}